\UseRawInputEncoding
\documentclass[]{article}

\usepackage{amsmath}
\usepackage{graphicx}
\usepackage{booktabs}
\usepackage{array}
\usepackage{hyperref}
\usepackage{multirow}
\usepackage{xcolor}
\title{Smoothed Model-Assisted Small Area Estimation}
\author{Peter A. Gao and Jon Wakefield}

\begin{document}

\maketitle

\begin{abstract}
In countries where population census data are limited, generating accurate subnational estimates of health and demographic indicators is challenging. Existing model-based geostatistical methods leverage covariate information and spatial smoothing to reduce the variability of estimates but often ignore survey design, while traditional small area estimation approaches may not incorporate both unit level covariate information and spatial smoothing in a design-consistent way. We propose a smoothed model-assisted estimator that accounts for survey design and leverages both unit level covariates and spatial smoothing. Under certain assumptions, this estimator is both design-consistent and model-consistent. We compare it with existing design-based and model-based estimators using real and simulated data.
\end{abstract}


\section{INTRODUCTION}\label{s:intro}

Subnational estimates of health and demographic indicators such as immunization coverage are critical for policy design and assessing inequality between regions. When census data is unavailable, household surveys can provide information at the national level, but may not be designed to produce reliable subnational estimates at the level required for decision-making, especially when estimating the prevalence of rare events. Estimates are often desired for ``unplanned" domains, meaning regions that do not align with the survey design, or for regions or subpopulations for which sample sizes are insufficient, called ``small areas." The problem of obtaining reliable estimates in this setting is called small area estimation and has motivated research at the intersection of survey statistics, hierarchical modeling, and spatial statistics. Pfeffermann (2013) and Rao and Molina (2015) review recent advances in small area estimation while Wakefield et al. (2020) consider small area estimation in the context of disease prevalence mapping. Small area estimation methods have also been used for subnational mapping of poverty indicators (Bell, Basel, and Maples, 2016; Marhuenda et al., 2017; Corral, Molina, and Nguyen, 2020), health outcomes (Congdon and Lloyd, 2010), and crop production estimates (Erciulescu, Franco, and Lahiri, 2019).

When response data are limited, ``direct" weighted estimators, which rely solely upon a given area's response data to estimate a corresponding area level quantity, can be unreliable with large sampling errors.  As a result, ``indirect" modeling methods that share information between areas can be preferable. Such models typically incorporate covariate information or smooth across areas using random effects and can produce more precise estimators than direct estimation. Small area models can be divided into two general categories: area level models and unit level models. Area level models model area specific quantities (such as survey weighted area means) and can incorporate area level covariates. When survey microdata are available, unit level models, which model individual responses, can be used.

Area level models are often applied to smooth direct weighted estimators which are typically design-consistent and asymptotically design unbiased. The resulting model-based estimators often inherit favorable design optimality properties and since only aggregate quantities are modeled, fewer assumptions about the distribution of individual responses values are needed. On the other hand, unit level models can incorporate higher resolution covariate information and are more easily adapted for modeling binary responses or count data.  For both area level models and unit level models, the default choice is to assume that any area- or cluster-level random effects are independently distributed, but models with spatially correlated random effects may improve precision when there is spatial structure in responses not explained by observed covariates (Chung and Datta, 2020). Although methods exist for survey weighted estimation of unit level models with independent random effects, the problem of achieving design-consistent small area estimators using unit level models with spatially correlated random effects is not well understood. 

In many low- and middle- income countries (LMIC), census data on highly informative unit level covariates may be limited, so it has become popular to use geostatistical unit level mixed models with satellite-observed covariates (such as nighttime light emissions) and spatial random effects when mapping health and demographic indicators. Such methods have been used to develop high-resolution maps of disease prevalence (Diggle, 2016), vaccination rates (Utazi et al., 2020), and neonatal and child mortality (Golding et al., 2017). However, such models often do not account for the complex design of surveys used in this setting. For example, the Demographic and Health Surveys (DHS) Program, which collects health outcomes data in many countries, typically uses a multistage stratified clustered design, oversampling clusters in urban areas.

When using unit level models that incorporate unit level covariates and spatial random effects, neglecting to carefully consider the survey design can lead to biased or poorly calibrated estimators. We are principally concerned with two potentially intertwined issues: informative sampling and clustering. Under informative sampling, where the sample response is correlated with the inclusion probability even after conditioning on model covariates, estimators derived from unit level mixed models may be biased unless the estimation procedure is adjusted to account for this dependence (Pfeffermann and Sverchkov, 2007; Parker, Janicki, and Holan, 2020).  Similarly, when cluster sampling is used, failing to account for within-cluster correlation may reduce the accuracy of point estimates and result in improperly calibrated interval estimates. When reliable sampling weights are available, they can be used when estimating model parameters (for example, via pseudo-likelihood or composite likelihood methods) to address these issues. However, many survey organizations including the DHS generally publish only single-inclusion sampling weights derived from each sampled individual's overall inclusion probability.  For multistage designs where sampling of clusters is informative, more detailed information on sampling probabilities at each stage may be needed to achieve design-consistent estimation for mixed model parameters, as shown by Slud (2020b).

To address these difficulties, we propose a two-stage smoothed model-assisted estimator that incorporates unit level covariate information and spatial smoothing. We draw from both area level and unit level methods by first using a working {\it unit level} model to generate area level predictions which we subsequently smooth by applying an {\it area level} model with spatial random effects. We only incorporate smoothing using random effects in the second stage, at the area level, allowing us to avoid some of the challenges associated with incorporating sampling weights when estimating parameters for unit level mixed models. Our method can be viewed as a bridge between traditional small area estimation approaches and geostatistical models commonly used in global health research which may not explicitly account for the survey design. We use recent advances in spatial modeling and outline a fully Bayesian approach to estimation and inference, showing how modern Bayesian methods and computationally efficient software commonly used in model-based geostatistics and global health research can be adopted for small area estimation.

The remainder of this paper details our proposed method and draws connections between our approach and existing methods. In Section \ref{s:motiv}, we introduce our motivating example of estimating the spatial distribution of measles vaccination rates in Nigeria based on DHS data. In Section \ref{s:existing}, we review existing small area estimation methods and discuss their applicability in our context. In Section \ref{s:methods}, we outline our smoothed model-assisted estimator.  We compare our estimator with existing methods in a simulation study in Section \ref{s:sims} and in the application of estimation of measles vaccination rates in Section \ref{s:mcv}. We discuss our method, offer practical advice for small area estimation in this context, and suggest future areas of research in Section \ref{s:disc}.

\section{MOTIVATION}\label{s:motiv}

In this section, we outline a motivating example: estimation of measles vaccination coverage for subnational areas in Nigeria using data from the Demographic and Health Surveys (DHS) Program, as previously described by (Fuglstad, Li, and Wakefield, 2021). We describe the importance of acknowledging the survey design and discuss how limited data availability makes subnational estimation difficult. The challenges encountered in this setting are characteristic of other estimation problems involving health and demographic indicators in LMIC.

In many LMIC, the DHS Program conducts regular surveys to collect data on population and health, including  vaccination coverage. The DHS Program generally uses a stratified two-stage cluster sampling design within each country. Countries are divided into principal administrative divisions, usually called Admin-1 regions. These Admin-1 regions are further divided into urban and rural divisions and sampling is stratified by strata obtained by crossing Admin-1 region with urban/rural status. Each strata is further divided into smaller collections of households called enumeration areas (EAs) or clusters. Using available census data, a specified number of EAs in each stratum is sampled with probability proportional to the number of households in the EA. Finally, the households in each selected EA are enumerated, and a specified number of households is sampled from each. Inclusion probabilities are calculated for each household and sampling weights can be subsequently derived. 

\begin{figure}
	\centering
	\includegraphics[scale=.45]{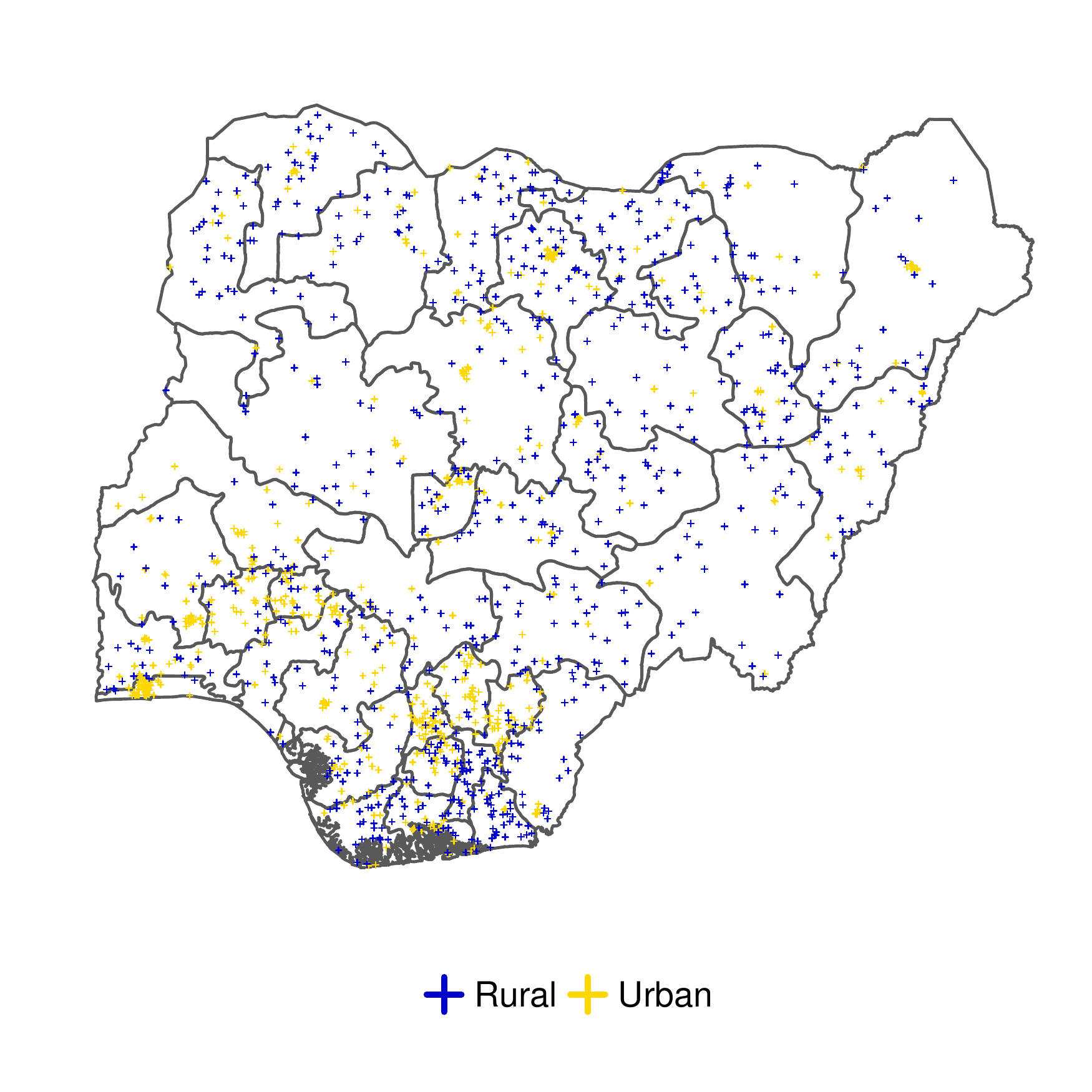}
	\caption{Map of Nigeria with Admin-1 level boundaries (thick borders) and Admin-2 level boundaries (thin borders). Points indicate enumeration area locations for which data on measles vaccination is available.}
	\label{fig:map}
\end{figure}

Our goal is to generate estimates of subnational vaccine coverage rates for the first dose of measles-containing-vaccine (MCV1) among children aged 12-23 months in Nigeria using data from the 2018 Nigeria DHS. The 2018 DHS collected data on vaccination status for children in sampled households based on vaccination cards or caregiver recall. We desire estimates at the Admin-1 level, which in Nigeria comprises 36 states and the Federal Capital Territory of Abuja, referring to the Database of Global Administrative Areas (GADM) boundaries (https://gadm.org/download\_country\_v3.html). The sampling frame used for the 2018 DHS was based on a national census conducted in 2006 which divided Nigeria into 664,999 EAs and 74 strata (obtained by splitting Admin-1 areas by urban/rural locations). Data were successfully collected in 1389 EAs; a number of clusters were dropped due to security issues during the household listing operation. As a result, as noted in Appendix A.3 of the Nigeria DHS Final Report,  estimates for the Admin-1 area of Borno may not be representative of omitted EAs.  Geographic coordinates are available for almost all EAs, but locations have been displaced by small distances to maintain privacy. Figure \ref{fig:map} provides a map of the Admin-1 boundaries and EA locations in Nigeria for which data is available.

The multi-stage stratified cluster design used by the DHS program complicates estimation of subnational means and totals. Urban EAs are oversampled relative to rural EAs, meaning that estimation methods must account for any systematic differences between urban and rural households. In the case of measles vaccination in Nigeria, urban areas exhibit higher rates of vaccination than rural areas (Fuglstad, Li, and Wakefield, 2021). As mentioned in Section \ref{s:intro}, In both the small area estimation and geostatistics literatures, modeling often assumes that the survey design is ignorable with respect to the model used for estimation, meaning that the same model applies to both sampled and non-sampled data. When this assumption fails, the resulting estimators may be biased. In addition, estimates based on clustered sample data can lead to underestimates of variability if models do not account for within-cluster correlation. As such, both small area estimation models and geostatistical models (and their estimation procedures) should be designed with consideration for the survey design.

Although DHS data are often adequate for computing reliable direct estimators of indicators at the Admin-1 level, sufficient data may not be available at subregional levels, where estimates for non-sampled regions are derived using data from other areas. Covariate information and smoothing via random effects can be used to generate estimates with reduced variability. 

\section{EXISTING APPROACHES}\label{s:existing}

\subsection{Notation}

We focus on estimation of area-specific proportions such as rates of vaccination coverage. Let $U=\{1,\ldots,N\}$ denote a set of indices for a finite population of size $N$. We assume $U$ is partitioned into $A$ disjoint administrative areas, $U=U_1\cup \cdots\cup U_A$, with $U_a$ being the set of $N_a$ indices corresponding to units in area $a$. For all $i\in U$, we use $y_i$ to denote the value of a variable of interest  for unit $i$. In our vaccination coverage example, $y_i\in\{0, 1\}$ are binary variables with a value of 1 indicating vaccination. Our targets of estimation are the area-specific vaccination coverage rates $p_a:$
\begin{equation}p_a=\frac{1}{N_a}\sum_{i\in U_a}y_i\label{e:target}\end{equation}

We use $S=\{j_1,\ldots, j_n\}\subset U$ to denote a random set of $n$ sampled indices, letting $S=S_1\cup\cdots\cup S_A$ be the corresponding partition by administrative area. We assume a probability sampling scheme, where $S$ is random, and for all $i\in U$, we let $\pi_i $ be the probability that $i\in S$, also called the inclusion probability of unit $i$. Finally, we let $w_i=1/\pi_i$ denote the sampling weight for unit $i$ defined as the inverse inclusion probability.

\subsection{Area level model-based estimation}

Direct weighted estimators for an area-specific mean use only data from the area in question. One standard direct estimator, the H\'ajek estimator  (H\'ajek, 1971) extends the Horvitz-Thompson estimator (Horvitz and Thompson, 1952), using sampling weights to approximate the totals in Equation (\ref{e:target}) :
\begin{equation}
\widehat{p}_a^{H}=\frac{\sum\limits_{i\in s_a}w_iy_i}{\sum\limits_{i\in s_a}w_i}
\end{equation}
For many sequences of designs and populations, $\widehat{p}_a^{H}$ is a design-consistent estimator of $p_a$ (for a definition of design consistency, see Appendix \ref{app:asym}). When data are limited, direct estimators such as $\widehat{p}_a^{H}$ can be unreliable and model-based methods may be used to leverage smoothing and auxiliary information. Area level models assume direct estimates are noisy observations of true area-specific quantities. The standard Fay-Herriot model combines a sampling model for the direct estimators with a linking model for the true finite population means $p_a$ (Fay and Herriot, 1979). We can specify the combined model as follows, using $\widehat{p}_a^H$ to denote the H\'ajek estimator:
\begin{align}
\widehat{p}_a^H &= p_a + \epsilon_a\label{e:fh-sampling}\\
p_a &= \mathbf{x}_a^T\boldsymbol\beta +u_a\label{e:fh-linking}
\end{align}
where for $a=1,\ldots A$, $u_a\stackrel{iid}{\sim} N(0, \sigma_u^2)$ are independently and identically distributed area-specific random effects that are also independent of sampling errors $\epsilon_a\sim N(0, V_a)$. We let $V_a$ denote the design variance of $\widehat{p}_a^H$ which is assumed known. In practice, $V_a$ is estimated or approximated. Finally, $\mathbf{x}_a$ represents a vector of area-specific covariates and $\boldsymbol\beta$ denotes the corresponding vector of coefficients.  Note that although we use $\widehat{p}_a^H$ as our direct estimator, we could replace it with another direct or design-consistent estimator. This basic area level Fay-Herriot model can be used to generate model-based estimates either by taking a frequentist approach and computing the empirical best linear unbiased predictor (EBLUP) or by using a Bayesian approach to sample from the posterior distribution of $p_a$; for more details see Chapters 6 and 9 of Rao and Molina (2015). Assuming design consistency of the direct estimator and a sequence of designs and populations such that $V_a\rightarrow 0$, the EBLUP is also a design-consistent estimator of $p_a$ (Section 6.1, Rao and Molina, 2015). 

The basic area level model assumes that area random effects $u_a$ are identically and independently distributed, but this model has been extended to allow for random effects with spatial and spatiotemporal correlation structures (Ghosh et al., 1998; Petrucci and Salvati, 2006; Pratesi and Salvati, 2008). Chung and Datta (2020) found that extending the Fay-Herriot model to use spatial models for the area effects $\mathbf{u}$ may improve upon the base model when there is spatial structure in the direct estimators not explained by observed area level covariates. When estimating demographic rates, covariate information is often derived from available censuses, but in many LMIC, such data may be limited. Incorporating spatial smoothing via random effects may help account for between-area differences related to unmeasured covariates. Mercer et al. (2015) use a Fay-Herriot type model of logit-transformed direct estimators with spatiotemporal random effects to estimate child mortality rates. In addition to spatial random effects modeling, Porter et al. (2014) extend the Fay-Herriot model to include functional covariates based on sources such as satellite imagery, which may be more readily available than typical census-based auxiliary information.

\subsection{Unit level model-based estimation}
When microdata or more detailed covariate information are available, unit level models, which directly model individual responses, can improve upon area level models (Hidiroglou and You, 2016). Such methods model covariate relationships at the individual response level instead of at an aggregate level. For continuous responses, the nested error regression model, also called the basic unit level model, was proposed by Battese, Harter, and Fuller (1988):
\begin{equation}\label{e:ulm}
y_i = \mathbf{z}_i^T\boldsymbol\gamma+u_{a(i)}+\varepsilon_i
\end{equation}
Above, $\mathbf{z}_i$ denotes covariate values for individual $i$ and $\boldsymbol\gamma$ denotes the corresponding coefficients. The area in which unit $i$ is located is denoted $a(i)$, with $u_{a(i)}\stackrel{iid}{\sim} N(0,\sigma_a^2)$ representing an area level random effect. Finally, $\varepsilon_i\stackrel{iid}{\sim} N(0,\sigma_\varepsilon^2)$ represent random and independent measurement error. The above model uses one level of random effects for each area, but for multistage designs, unit level models could also include random effects for each stage of sampling, as suggested by Marhuenda et al. (2017). Predictions for a particular area can be generated by aggregating unit level predictions.

For a binary response, a binomial unit level mixed model can be specified:
\begin{equation}\label{e:logistic-ulm}
y_i\mid \mathbf{z}_i, \boldsymbol\gamma, u_{a(i)}\sim \mathrm{Binomial}(1, q_i)
\end{equation}
\begin{equation}\label{e:logistic-ulm2}
\mathrm{logit}(q_i) = \mathbf{z}_i^T\boldsymbol\gamma+u_{a(i)}
\end{equation}
where $q_i$ denotes a individual level risk parameter. For other types of response data, other likelihoods may also be used for modeling unit level response data.

While estimators based on basic area level models are often design-consistent, unit level models may not generally produce design-consistent estimators. In particular, as mentioned above, informative sampling or cluster sampling may complicate estimation using unit level models. Parker, Janicki, and Holan (2020) provide an excellent and comprehensive overview of existing strategies used to address the design when using unit level model. Below, we comment on some of the most popular methods for incorporating sampling weights when estimating model parameters.

Pseudo-likelihood methods, as described by Binder (1983) and Skinner (1989) incorporate survey weights into model estimation of linear and generalized linear models, as reviewed by Lumley and Scott (2017).   This has been extended to estimation for both linear mixed models (Pfeffermann et al., 1998)  and generalized linear mixed models (Asparouhov, 2006; Rabe-Hesketh and Skrondal, 2006) with multiple levels of random effects for multistage sampling designs. In practice, separate weights may be needed to account for cluster level effects and unit level effects, but many surveys, including those used by the DHS, only provide one set of sampling weights corresponding to the final inclusion probabilities of each unit. As noted by Slud (2020b), under informative multistage sampling, mixed model parameters may be nonidentifiable if only final single inclusion weights are available. In addition, using unscaled weights can lead to bias in estimation of variance parameters and subsequently small area means (Pfeffermann et al., 1998; Korn and Graubard, 2003; Asparouhov, 2006). Other alternatives include estimation via pairwise likelihood using pairwise inclusion probabilities (Rao, Verret, and Hidiroglou, 2013; Yi, Rao, and Li, 2016) or direct modeling of survey weights (Pfeffermann and Sverchkov, 2007). All of these methods acknowledge the design, but may be sensitive to scaling of weights or require availability of higher-order or pairwise sampling weights. 

Unit level models like those in Equation (\ref{e:ulm}) have also been extended to account for spatial variation. Chandra, Salvati, and Chambers (2007) examine the use of a nested error regression model with spatial area effects and Chandra et al. (2012) use a geographically weighted regression approach that allows fixed effects coefficients to vary across all possible cluster locations. However, the resulting estimators do not use survey weights and the effect of informative sampling upon parameter estimation is not clear. Huang (2019) outlines an approach using pairwise likelihood to estimate spatial correlation parameters for a cluster level model but it requires knowledge of pairwise sampling probabilities. 

In recent years, global health researchers have also turned to unit level modeling using satellite-derived covariate information and spatial random effects modeled using latent spatial Gaussian processes for mapping health indicators in LMIC. This body of research often does not refer heavily to the small area estimation literature, instead arising out of research in geostatistics. Typically the sampling design is assumed to be ignorable with respect to the models used, which do not directly acknowledge the survey design using survey weights. In these unit level models, the latent Gaussian processes are often assumed to vary smoothly in space, enabling researchers to produces maps of health indicators at resolutions as fine as 1 km by 1 km (Diggle and Giorgi, 2019; Utazi et al., 2020). When the number of prediction locations is high, using Gaussian process models can be time-consuming, but approximate methods can speed up computation.  The integrated nested Laplace approximation-stochastic partial differential equation (INLA-SPDE) approach is popular for approximate Bayesian inference with spatial and spatiotemporal Gaussian process models (Rue, Martino, and Chopin, 2009; Lindgren, Rue and Lindstr\"om, 2011). Although the continuous spatial modeling approach allows for prediction at any location, interpretation of the continuous surface is complicated: the surface is assumed to exist even at locations where no individual is present. For this reason, it may be preferable to model actual prevalences among groups of individuals in the finite population.

Such models can adjust for clustering by incorporating cluster level random effects or adopting alternative likelihoods that account for overdispersion; however, without complete census frame information, it is not obvious how to aggregate cluster effects when generating predictions. Furthermore, geostatistical models may not always explicitly account for urban/rural stratification when using data from the DHS and other surveys. Paige et al. (2020) and Dong and Wakefield (2021) have shown that urbanicity can be associated with health outcomes, leading to bias if the stratification is not incorporated into the model. An additional complication results from changing levels of urbanization over time. From a design-based perspective, it is important to account for the urban/rural stratification used at the time of sampling as a unit's inclusion probability depends on its sampling stratum. However, in many LMIC, increasing urbanization means that clusters in rural strata may change over time to resemble urban clusters more closely.  Geostatistical models often incorporate covariates like intensity of night time lights that could be viewed as surrogates for urbanicity, but such covariates may not align with the original partition used to define sampling strata.

\section{SMOOTHED MODEL-ASSISTED ESTIMATION}\label{s:methods}

Below, we present a small area estimator that accounts for survey design while incorporating unit level covariate information and smoothing via random effects. In essence, our two-stage smoothed model-assisted method uses an area level model to smooth estimates obtained via a model-assisted approach. We first use a working model with unit level covariates to calculate a generalized regression estimator and then use a linking model to induce smoothing on the model-assisted estimators. As long as the model-assisted estimators are design-consistent and their design variances converge to zero, our smoothed model-assisted estimators will also be design-consistent. We thus provide an alternative to existing small area estimation methods that only requires final sampling weights. Our proposed model incorporates unit level covariates and spatial random effects while also explicitly accounting for the sampling design.

\subsection{Stage One: Model-Assisted Estimation}\label{ss:MAE}
While the unit level models described in Section \ref{s:existing} may produce biased estimators under model misspecification, model-assisted estimators are motivated by working superpopulation models but are specified to ensure design consistency and unbiasedness even when the working model is incorrect. For a review of model-assisted methods see S\"arndal et al. (2003) or Breidt and Opsomer (2017). A characteristic model-assisted approach is given by the difference estimator:
\begin{equation}
\widehat{p}_a^{DIFF}=\frac{1}{N_a}\left\{\sum_{i\in U_a} \widehat{y}_i+\sum_{i\in s_a}w_i(y_i-\widehat{y}_i)\right\}
\label{e:diff}
\end{equation}
where $\widehat{y}_i$ represents the working model prediction for unit $i$. The difference estimator for area $a$ combines model-based predictions from the working model with a direct estimator of the mean of the residuals in the area based upon the sample. Breidt and Opsomer (2017) show that under certain regularity conditions, the difference estimator is design-consistent. In particular, they assume the direct estimator for the residual mean is design-consistent and that predictions from the working model estimated on sample data and predictions from the working model estimated on the full population are asymptotically equivalent. The popular generalized regression estimator (GREG) can be framed as an example of a difference estimator using a working linear regression model to generate predictions (S\"arndal et al., 2003). In the case of binary response data, Lehtonen and Veijanen (1998) previously proposed the use of a working logistic regression model to compute a logistic generalized regression (LGREG) estimator. Kennel and Valliant (2010) extended the LGREG for use with cluster sample and Myrskyla (2007) compared the LGREG and GREG with binary responses, finding that when the model fit is strong, the LGREG is preferable. 

For the first stage of our smoothed model-assisted approach for estimating small area proportions, we compute a model-assisted estimator using a working logistic regression model of the form:
\begin{align}
P(y_i=1\mid \mathbf{z}_i, \boldsymbol\gamma)&=q_i\label{e:logistic-wm}\\
\mathrm{logit}(q_i) &= \mathbf{z}_i^T\boldsymbol\gamma\label{e:logistic-wm2}
\end{align}

For the basic GREG estimator, the working model is generally fit separately for each area, but in order to obtain more stable parameter estimates, we use a global model for all areas. In this sense, our approach resembles the ``modified GREG" estimator described by Rao and Molina (2015), also referred to as the survey regression estimator. We estimate model parameters via survey-weighted maximum likelihood and then generate working predictions $\widehat{y}_i=\mathrm{expit}(\mathbf{z}_i^T\widehat{\boldsymbol\gamma})$ for all $i\in U$.

Based on these working predictions, we construct the model-assisted estimator:
\begin{equation}\label{e:greg-h}
\widehat{p}_a^{MA}=\frac{1}{\widehat{N}_a}\left(\sum_{i\in U_a}\widehat{y}_i+\sum_{i\in S_a}w_i(y_i-\widehat{y}_i)\right)
\end{equation}
where $\widehat{N}_a=\sum_{i\in S_a}w_i$, yielding a H\'ajek-like estimator. Under certain regularity conditions, this estimator is design-consistent; for further details see Appendix \ref{a:proof}.  

Model-assisted estimators are typically asymptotically design unbiased and design-consistent, but quantification of uncertainty can be difficult. Linearization-based variance approximations generally do not account for uncertainty in the first sum on the right of Equation (\ref{e:diff}) resulting from model estimation (Myrskyla, 2007). The working model should be carefully selected as overfitting can also result in underestimation of uncertainty. For our model-assisted estimator, we estimate variance by modifying the with-replacement variance estimator of a total described by Kennel and Valliant (2010) for use with a mean:
\begin{equation}\label{e:V-MA}
\widehat{V}(\widehat{p}_a^{MA})=\frac{1}{\widehat{N}_a^2}\frac{n_a}{(n_a-1)}\sum_{i\in S_a}\left(w_ie_i-\widehat{\overline{e}}_a\right)^2
\end{equation}
where $n_a$ denotes sample size for area $a$, $\widehat{\overline{e}}_a= \frac{1}{n_a}\sum_{i\in S_a}w_ie_i$, and $e_i=\widehat{y}_i-y_i$. This estimator is designed for unclustered sampling designs; when applying our approach to DHS data, we adapt Kennel and Valliant's cluster sampling variance estimator. Note that this variance estimator ignores variability resulting from $\widehat{N}_a$ and estimation of the regression parameters.  In practice, variance estimation may be improved via resampling methods such as the bootstrap.

\subsection{Stage Two: Spatial Logistic Area Level Model}\label{ss:slm}

After computing the model-assisted estimators and their associated variance estimators, we use a Fay-Herriot model to smooth across areas. Since our targets of estimation $p_a$ are bounded between 0 and 1, we incorporate a logit transformation into both the sampling and linking models. In essence, we apply a spatial area level model to logit-transformed model-assisted estimators. Our linking and sampling models can be specified as follows: 
\begin{align}
\mathrm{logit}(p_a) &= \mathbf{x}_a^T\boldsymbol\beta + u_a\label{e:slm-linking}\\
\mathrm{logit}(\widehat{p}_a^{MA})&= \mathrm{logit}(p_a) + \epsilon_a\label{e:slm-sampling}
\end{align}
where for $a=1,\ldots, A$, $\mathbf{x}_a=(1, x_{a1}, \ldots, x_{ap})^T$ represents a length $p+1$ vector of area-specific covariates and $\boldsymbol\beta=(\beta_0, \beta_1, \ldots, \beta_p)^T$ denotes the vector containing the intercept and corresponding fixed effect coefficients. We use $\mathbf{u}=(u_1,\ldots, u_A)^T$ to denote random area level effects, which we assume to be spatially correlated and drawn from a multivariate Gaussian distribution,  $\mathbf{u}\sim \boldsymbol{\mathcal{N}}(\mathbf{0},\boldsymbol\Sigma( \sigma_u^2, \phi))$. Here, $\sigma_u$ and $\phi$ denote parameters controlling the spatial correlation matrix $\boldsymbol\Sigma$. Finally, we use $\epsilon_a$ to denote independent sampling errors $\epsilon_a\sim N(0, V_a)$, where $V_a=\mathrm{Var}(\mathrm{logit}(\widehat{p}_a^{MA}))$, which we treat as known. In practice, we estimate $V_a$ by first estimating $\widehat{V}(\widehat{p}_a^{MA})$ using Equation (\ref{e:V-MA}) and then applying the delta method to obtain the approximation:
\begin{equation}
V_a\approx \frac{\widehat{V}(\widehat{p}_a^{MA})}{(\widehat{p}_a^{MA}(1-\widehat{p}_a^{MA}))^2}
\end{equation}
We adopt a hierarchical Bayesian approach to inference by defining hyperparameter priors, yielding the following alternative representation:
\begin{align}
\mathrm{logit}(\widehat{p}_a^{MA}) \mid p_a,  V_a &\stackrel{ind}{\sim}  N(\mathrm{logit}(p_a), V_a),&a =1,\ldots A\\
\mathrm{logit}(\mathbf{p}_a) \mid \boldsymbol\beta,  \sigma_u^2, \phi &\sim \boldsymbol{\mathcal{N}}(\mathbf{X}\boldsymbol\beta, \boldsymbol\Sigma( \sigma_u^2, \phi))\\
\boldsymbol\beta, \sigma_u^2, \phi &\sim \pi(\boldsymbol\xi)
\end{align}
where we use $\mathrm{logit}(\mathbf{p}_a)$ to denote the vector $(\mathrm{logit}(p_1), \ldots, \mathrm{logit}(p_A))^T$, and $\mathbf{X}$ to denote the $A\times (p+1)$ matrix of area level covariates. Finally, $\pi(\boldsymbol\xi)$ denotes the hyperparameter priors and $\xi$ represents corresponding parameters, which should be specified based on the specific application at hand. 

By specifying different structures for $\boldsymbol\Sigma$, we can obtain varying models for the spatial dependence in $\mathrm{logit}(\mathbf{p}_a)$. Typically, we specify an $A\times A$ adjacency matrix representing adjacency relationships between the areas. We model the area level random effects $\mathbf{u}$ using the BYM2 model, a reparametrization of the Besag-York-Molli\'e (Besag, York, and Molli\'e, 1991) model proposed by Riebler et al. (2016) for the area level random effects vector $\mathbf{u}$:
\begin{equation}
\mathbf{u}=\sigma_u\left(\sqrt{1-\phi}\tilde{\mathbf{u}}_1+\sqrt{\phi}\tilde{\mathbf{u}}_{2*}\right)
\end{equation}
Here, we assume $\tilde{\mathbf{u}}_1\sim N(\mathbf{0}, \mathbf{I})$ is an random area effect with no spatial structure. We use $\tilde{\mathbf{u}}_{2*}$ to denote a structured spatial component which follows an intrinsic conditional autoregressive (ICAR) model (intuitively, the mean of $\tilde{u}_{2i*}$ is set to the mean of all neighboring effects and the precision is specified to be proportional to the number of neighbors). As such, $\sigma_u$ controls the marginal variance of $\mathbf{u}$ and $\phi$ controls the proportion of variation assigned to the structured component. 
Under this model, $\mathbf{u}$ has the covariance matrix
\begin{equation}
\mathrm{Var}(\mathbf{u}\mid \sigma_u, \phi)=\sigma_u((1-\phi)\mathbf{I}+\phi \mathbf{Q}_{*}^{-})
\end{equation}
Above, $\mathbf{Q}_{*}^{-}$ is the generalized inverse of $\mathbf{Q}_{*}$, which denotes the precision matrix of $\tilde{\mathbf{u}}_{2*}$. As discussed by Riebler et al. (2016), $\mathbf{Q}_{*}$ is singular, making the ICAR prior for the random effects improper, so we place a sum to zero constraint on the elements of $\mathbf{u}$ to ensure identifiability. Moreover, the marginal variance of each effect $\mathbf{u}_a$ depends on its number of neighbors, so to make the overall variance parameter $\sigma_u$ interpretable,  $\mathbf{Q}_{*}$ is scaled (following the procedure described in Section 3.2 of Riebler et al. (2016)) to make the geometric mean of the marginal variances is equal to one.

Following Riebler et al. (2016) and Simpson et al. (2017), we place penalized complexity (PC) priors on $\sigma_u$ and $\phi$. These priors penalize the Kullback-Leibler distance of a full model from a simpler base model and shrink $\phi$ and $\sigma_u$ to zero. We place a flat prior on $\boldsymbol\beta$, so that $\pi(\boldsymbol\beta)\propto 1$. To fit our spatial logistic area level model, we use the R package \texttt{INLA} (Rue et al., 2017), which is commonly used to conduct approximate Bayesian inference for hierarchical models and is popular for mapping health indicators (Golding et al., 2017; Utazi et al., 2020).

The approach presented here is specialized for spatially structured binary response data, but the overall strategy of using area level models to smooth model-assisted estimators can be adapted for other types of data. If the response is continuous, rather than binary, the same approach can be applied using a working linear regression model in the first stage and using a similar second stage model, but without applying the logit transformations, as proposed by Fay (2018). Other models for spatial random effects could be used, including, for example, those previously introduced into Fay-Herriot models by Ghosh et al. (1998) and Pratesi and Salvati (2008), As mentioned above, Chung and Datta (2020) provide a recent comparison of several spatial random effects models used within the Fay-Herriot model. Examples of previous uses of ICAR spatial random effects within area level models are provided in You and Zhou (2011),  Porter et al. (2014), and Mercer et al. (2015).

\section{SIMULATIONS}\label{s:sims}

\subsection{Population generating model}

Below, we use simulations to compare our smoothed model-assisted estimator with existing direct, model-assisted, and model-based estimators. The set up is motivated by simulations used by Corral et al. (Section 7.2, 2020). Using the WorldPop 100m population counts grid for Nigeria corresponding to the 2006 census (Worldpop, 2006), we sample 300 pixels without replacement with probability proportional to population in each of 73 strata defined by crossing the 37 Admin-1 areas with urban/rural status (one area corresponding to Lagos is entirely urban). Each sampled pixel represents a simulated cluster location. We then randomly generate cluster sizes for each simulated cluster so that the size of cluster $c$ is given by $n_c\sim \mathrm{Poisson(15)}$.  For each cluster $c$, we simulate a cluster level risk $q_c$ using the model:
\begin{multline}
\mathrm{logit}(q_c)=x_{1,c} - x_{2,c} + 0.5x_{3,c} + 0.25x_{4,c} + 0.25x_{5,c} \\+ 1.5x_{6, c} + 0.1x_{7, c} + 0.1x_{8, c} + u_{a(i)}+v_{c}
\end{multline}
where $u_{a(i)}\stackrel{iid}{\sim} N(0,0.1^2)$ are independent and identically distributed area level random effects, and $v_c\stackrel{iid}{\sim} N(0,0.5^2)$ represents random and independent and identically distributed cluster level effects. The covariates are specified as follows:
\begin{enumerate}
	\item The covariate $x_{1,c}$ is the realized value of a binary random variable $X_{1,c}$ with $P(X_{1,c} = 1) = 0.5$;
	\item The covariate $x_{2,c}$ is the realized value of a binary random variable $X_{2,c}$ with $P(X_{2,c} = 1) = 0.3 + 0.5\frac{a(c)}{37}$;
	\item The covariate $x_{3,c}=x_{3,a(c)}$ is obtained from a $37\times 1$ random vector modeled as an ICAR random effect with marginal variance 1 for the Admin-1 areas.
	\item The covariate $x_{4,c}=x_{4,a(c)}$ is obtained from a $774\times 1$ random vector modeled as an ICAR random effect with marginal variance 1 for the Admin-2 areas.
	\item The covariate $x_{5,c}$ is obtained from a random vector generated using a stochastic partial differential equation (SPDE) -based approximation to a Gaussian process with Mat\'ern covariance (smoothness 1) and marginal variance of 1.
	\item The covariate $x_{6,c}$ is obtained from a random vector generated using an SPDE-based approximation to a Gaussian process with Mat\'ern covariance (smoothness 1) and marginal variance of 1.
	\item The covariate $x_{7,c}$ denotes estimated travel times to cities in 2015 (Weiss et al., 2018).
	\item The covariate $x_{8,c}$ denotes proportion of people per grid square living in poverty in 2010 (Tatem et al., 2017).
\end{enumerate} 
The covariates $\mathbf{x}_1$ and $\mathbf{x}_2$ represent informative non-spatial covariates, while $\mathbf{x}_3$, $\mathbf{x}_4$, $\mathbf{x}_5$, and $_6$ exhibit spatial correlation. The covariates $\mathbf{x}_7$ and $\mathbf{x}_8$ are based on real covariates commonly used for modeling health outcomes in LMIC. Based on the above cluster level risks, we generate responses $Y_i\sim\mathrm{Bernoulli}(q_{c(i)})$. As described above, our population consists of 300 clusters of varying sizes. From this population, we repeatedly sample ten clusters from each stratum, using all response values from each sampled cluster. We use an informative sampling scheme in which we oversample clusters with large values for $x_{6, c}$: clusters with values of $x_{6, c}$ in the top quartile are three times as likely to be sampled as clusters in the bottom three quartiles. Since the values of $\mathbf{x}_6$ are spatially correlated, this may induce spatial structure in the model residuals if this oversampling is not addressed when estimating model parameters. Based on this design, we compute sampling probabilities and design weights $w_i$ for each individual.  In practice, we generate the covariate values and cluster sizes once and then sample a list of indices identifying the sampled clusters. These indices and cluster characteristics are held constant across simulations but the response variables, area effects, and cluster effects are repeatedly regenerated.

\subsection{Estimation procedure}

For each simulation, we compute true population Admin-1 area level proportions $p_a$ and compare with several estimators computed from the sampled data. For all estimators that rely on covariate modeling, we consider two potential models, a reduced model and a full model. The full model includes all covariates except $\mathbf{x}_{4}$. We remove the area-specific covariate $\mathbf{x}_{4}$ in order to induce spatial correlation in the model residuals.  The reduced model includes all covariates except $\mathbf{x}_{4}$ and $\mathbf{x}_{6}$ so it does not account for the effect of oversampling the stratum defined by $\{x_{6,i}> \mathrm{median}(x_{6,i})\}$, meaning the design is not ignorable after conditioning on model covariates. Conversely, the full model partially accounts for this by including $\mathbf{x}_6$ as a covariate. Furthermore, for all model-based approaches incorporating smoothing via random effects, we consider both non-spatial smoothing using iid Gaussian area level random effects and spatial smoothing using the BYM2 model for area level random effects.

Below, we describe the estimators used for comparison. First, we compute the direct weighted \textbf{H\'ajek} estimator. We also compute model-assisted estimators (\textbf{MA}) using both the full and reduced models. Next, we compute several area level model-based estimators. Applying the spatial logistic area level model described in Section \ref{ss:slm} to the H\'ajek estimator yields a spatial smoothed H\'ajek (\textbf{SH}) estimator. Similarly, by applying the same model to the model-assisted estimator, we obtain our proposed spatial smoothed model-assisted estimator  (\textbf{SMA}). For comparison, we also compute non-spatial versions of smoothed estimators by assuming independent and identically distributed Gaussian random effects in the logistic area level linking model given in Equation (\ref{e:slm-linking}). 

Finally, we compute a number of unit level model-based estimators, using a \textbf{Binomial} model as well as two models designed to account for effects of clustering: a \textbf{Betabinomial} model and a lognormal-binomial \textbf{Lono-Binomial} model (Dong and Wakefield, 2021). These particular likelihoods, as used in small area estimation, are discussed in further detail by Dong and Wakefield (2021),, but we briefly outline their use here. First, we implement the binomial unit level model specified in Equations (\ref{e:logistic-ulm}) and (\ref{e:logistic-ulm2}). The betabinomial model accounts for overdispersion in our response data potentially related to clustering and can be specified as follows. Letting $c(i)$ denote the cluster of unit $i$, we assume each unit in a given cluster has the same risk $p_i=p_{c(i)}$:
\begin{align}
Y_i\mid q_i&\sim\mathrm{Bernoulli}(q_i)\\
q_i = q_{c(i)}\mid \mu_{c(i)},d&\sim \mathrm{Beta}(\mu_{c(i)}, d)\\
\mathrm{logit}(q_i) &= \mathbf{x}_i\boldsymbol\beta + u_a
\end{align}
where we parameterize the beta distribution via
\begin{align}
E(p_{c(i)}\mid \mu_{c(i)},d)&= \mu_{c(i)}\\
\mathrm{Var}(q_{c(i)}\mid \mu_{c(i)},d&= \frac{\mu_{c(i)}(1-\mu_{c(i)})}{d+1}
\end{align}
Above, $d$ denotes a dispersion parameter. The lognormal-binomial model (referred to by Dong and Wakefield as the Lono-Binomial Overdispersion model) instead assumes that
\begin{align}
y_i \mid \mathbf{z}_i, \boldsymbol\gamma, u_{a(i)}&\sim \mathrm{Bernoulli}(r_{c(i)})\label{e:lnlogistic-ulm}\\
\mathrm{logit}(r_i) = q_i +v_{c(i)}&=\mathbf{z}_i^T\boldsymbol\gamma+u_{a(i)}+v_{c(i)}\label{e:lnlogistic-ulm2}
\end{align}
where for all clusters $c$, $r_c$ denotes a cluster level parameter defined as the sum of the cluster level prevalence $q_c$ and iid Gaussian cluster level error $v_{c}$. For all of these models, we implement both non-spatial iid and spatial BYM2 models for the area level random effects $\mathbf{u}_a$. For all unit level models, area level estimates are made by making predictions of $q_c$ for all clusters in the population and then aggregating upwards to the area level.

Additional information on the estimation procedures, including information on software used and priors for model hyperparameters, can be found in the Appendix. Code for the simulations (and for the application detailed below) can be found on GitHub.

\subsection{Results}
For each method, we compute point estimates $\widehat{p}_a$ as well as 90\% interval estimates $(\widehat{p}_a^-,\widehat{p}_a^+)$. For each vector of estimates $\widehat{p}_a$, we compute root mean squared error (RMSE) and mean absolute error (MAE). We also compute the coverage of the 90\% interval estimates and the mean interval lengths (MIL) across all areas.
\begin{align}
\mathrm{RMSE}(\widehat{p}_a)&=\sqrt{\frac{1}{A}\sum_{a}(p_a-\widehat{p}_a)^2}\\
\mathrm{MAE}(\widehat{p}_a)&=\frac{1}{A}\sum_{a}|p_a-\widehat{p}_a|\\
\mathrm{Cov}_{90}(\widehat{p}_a)&=\frac{1}{A}\sum_{a}\mathbf{1}\{p_a\in(\widehat{p}_a^-,\widehat{p}_a^+)\}\\
\mathrm{MIL}_{90}(\widehat{p}_a)&=\frac{1}{A}\sum_{a}(\widehat{p}_a^+ - \widehat{p}_a^-)
\end{align}

\begin{table}
	\centering
	\scriptsize
	
\begin{tabular}{rlrrrr}
	\scriptsize

	& \textbf{Method} & \textbf{RMSE} & \textbf{MAE} & \textbf{90\% Cov.}&\textbf{MIL}\\\hline
	&Direct (Hájek) & 4.44 & 3.28 & 86 & 13.99\\
	&MA & 3.70 & 2.81 & 87 & 11.82\\
	\hline
	\multirow{5}{*}{\textit{Non-spatial}}&SH& 4.84 & 3.44 & 89 & 14.66\\
	&SMA & 3.71 & 2.79 & 87 & 12.27\\
	&Binomial & 4.22 & 3.11 & 75 & 8.24\\
	&Betabinomial & 4.13 & 3.05 & 83 & 10.10 \\
	&Lono-Binomial & 4.42 & 3.23 & 81 & 9.95\\
	\hline
	\multirow{5}{*}{\textit{Spatial}}&SH  & 4.24 & 3.13 & 91 & 13.40\\
	&SMA & \textbf{\textit{3.47}} & \textbf{\textit{2.65} }& 87 & 11.49\\
	&Binomial & 4.14 & 3.02 & 76 & 8.06\\
	&Betabinomial & 3.96 & 2.88 & 85 & 9.84\\
	&Lono-Binomial & 4.38 & 3.18 & 81 & 9.72\\\hline
\end{tabular}
	\caption{Averaged RMSE ($\times 100$), MAE ($\times 100$), coverage rates, and MIL ($\times 100$) of estimators of area level means across 1,000 simulated populations with spatially correlated binary responses based on sample data obtained via informative sampling for methods using no covariates or only the reduced set of covariates (omitting one of the spatial covariates used in population generation). The lowest RMSE and MAE are in \textbf{\textit{bold italics}}.}
	\label{tab:sp-inf-model}
\end{table}
\begin{table}
	\centering
	\scriptsize
	
\begin{tabular}{rlrrrr}
	\hline
	\scriptsize
	& \textbf{Method} & \textbf{RMSE} & \textbf{MAE} & \textbf{90\% Cov.}&\textbf{MIL}\\\hline
	&MA & 3.17 & 2.41 & 87 & 9.68\\
	\hline
	\multirow{4}{*}{\textit{Non-spatial}}&SMA & 3.18 & 2.41 & 87 & 9.92\\
	&Binomial & 2.68 & 2.02 & 88 & 7.99\\
	&Betabinomial  & 2.69 & 2.04 & 89 & 8.43\\
	&Lono-Binomial& 4.42 & 3.23 & 81 & 9.95\\
	\hline
	\multirow{4}{*}{\textit{Spatial}}&SMA & 3.03 & 2.28 & 88 & 9.47\\
	&Binomial & \textbf{\textit{2.62}} & \textbf{\textit{1.96 }}& 88 & 7.76\\
	&Betabinomial & \textbf{\textit{2.62}} & \textbf{\textit{1.96 }}& 90 & 8.21\\
	&LobonBinomial & \textbf{\textit{2.62}} & \textbf{\textit{1.96 }}& 89 & 8.10\\\hline
\end{tabular}
	\caption{Averaged RMSE ($\times 100$), MAE ($\times 100$), coverage rates, and MIL ($\times 100$) of estimators of area level means across 1,000 simulated populations with spatially correlated binary responses based on sample data obtained via informative sampling for methods using the full set of covariates. The lowest RMSE and MAE are in \textbf{\textit{bold italics}}.}
	\label{tab:sp-inf-model-full}
\end{table}
We summarize these error metrics by averaging their values across all 1,000 simulated populations. For all estimators incorporating covariate information, we provide comparisons for both reduced (Table \ref{tab:sp-inf-model}) and full models (Table \ref{tab:sp-inf-model-full}) across the 1,000 generated response vectors. Note that the H\'ajek and SH estimators do not make use of any covariates. In general, introducing covariate information reduces the error of point estimates and the methods using the full set of covariates achieving the lowest error. The H\'ajek, model-assisted, and area level model-based estimators have coverage rates close to the nominal 90\% rate. However, some of the unit level model-based interval estimates exhibit undercoverage, especially when only the reduced set of covariates is used. For the full set of covariates, the estimators based on the binomial display undercoverage, while the betabinomial and lognormal-binomial interval estimates achieve close-to-nominal coverage. As expected, the spatial estimators generally improve upon their non-spatial versions.

\section{APPLICATION: VACCINATION COVERAGE IN NIGERIA}\label{s:mcv}

We apply our smoothed model-assisted estimator to generate Admin-1 level estimates of measles vaccination rates using the 2018 Nigeria DHS data described in Section \ref{s:motiv}. We use two main unit level covariates obtained from grid-based estimates of travel times to cities in 2015 (Weiss et al., 2018) and the proportion of people per grid square living in poverty in 2010 (Tatem et al., 2017). The associated fixed effect estimates were significantly different from zero in a survey-weighted logistic regression with measles vaccination as outcome; however, these covariates are themselves estimated using geostatistical models, so any associations should be interpreted with caution. We also use a map of estimated population density (WorldPop) to derive a binary covariate that classifies each pixel as either urban or rural. Below, $y_i$ denotes observed vaccination status of child $i$ and $\mathbf{x}_i$ denotes the corresponding covariate values.

When using unit level covariates to predict binary response variables, covariate information on the entire population is required to generate estimates. In our setting, when recent and reliable population data may not be available, satellite imagery can provide covariates on a pixel grid spanning the domain. Instead of predicting each child separately, we generate predictions for each pixel and average over the pixel level predictions for a given area.  When averaging, we weight each pixel's prediction by the estimated number of children aged 1-5 in the pixel using maps created by WorldPop We harmonize the covariate rasters and population density rasters to a common pixel grid, the 1km by 1km grid provided by WorldPop. We also use the map of estimated population density to derive a binary covariate that classifies each pixel as either urban  or rural assigning the highest density pixels in a given area to be urban so that the total proportion of population classified as urban in each area matches the proportion reported in the 2018 Nigeria DHS report.

Using this data, we compare a number of the estimation methods outlined above. We first consider the H\'ajek estimator $\widehat p_a^{H}$ and the model-assisted estimator $\widehat p_a^{MA}$ where the working model is a logistic regression model. We then consider smoothed H\'ajek estimators and smoothed model-assisted estimators obtained by fitting the model specified in (\ref{e:slm-linking}) and (\ref{e:slm-sampling}) for $\widehat p_a^{H}$ and $\widehat p_a^{MA}$, respectively. For each, we consider both iid and BYM2 models for the area level random effects $\mathbf{v}$, yielding four estimators: smoothed H\'ajek with iid area effects (\textbf{SH}) and BYM2 area effects (\textbf{Spatial SH}) as well as smoothed model-assisted with iid area effects (\textbf{SMA}) and BYM2 area effects (\textbf{Spatial SMA}). 

Finally, we consider a geostatistical model; to account for clustering, we use the \textbf{Spatial Betabinomial} model described above with a BYM2 prior for the area level random effects $\mathbf{u}$. Since pixels do not necessarily coincide with clusters, we cannot use the Lono-Binomial model, which requires us to identify the sampling frame of clusters in order to aggregate estimates appropriately. 

\begin{figure}
	\centering
	\includegraphics[scale=.43]{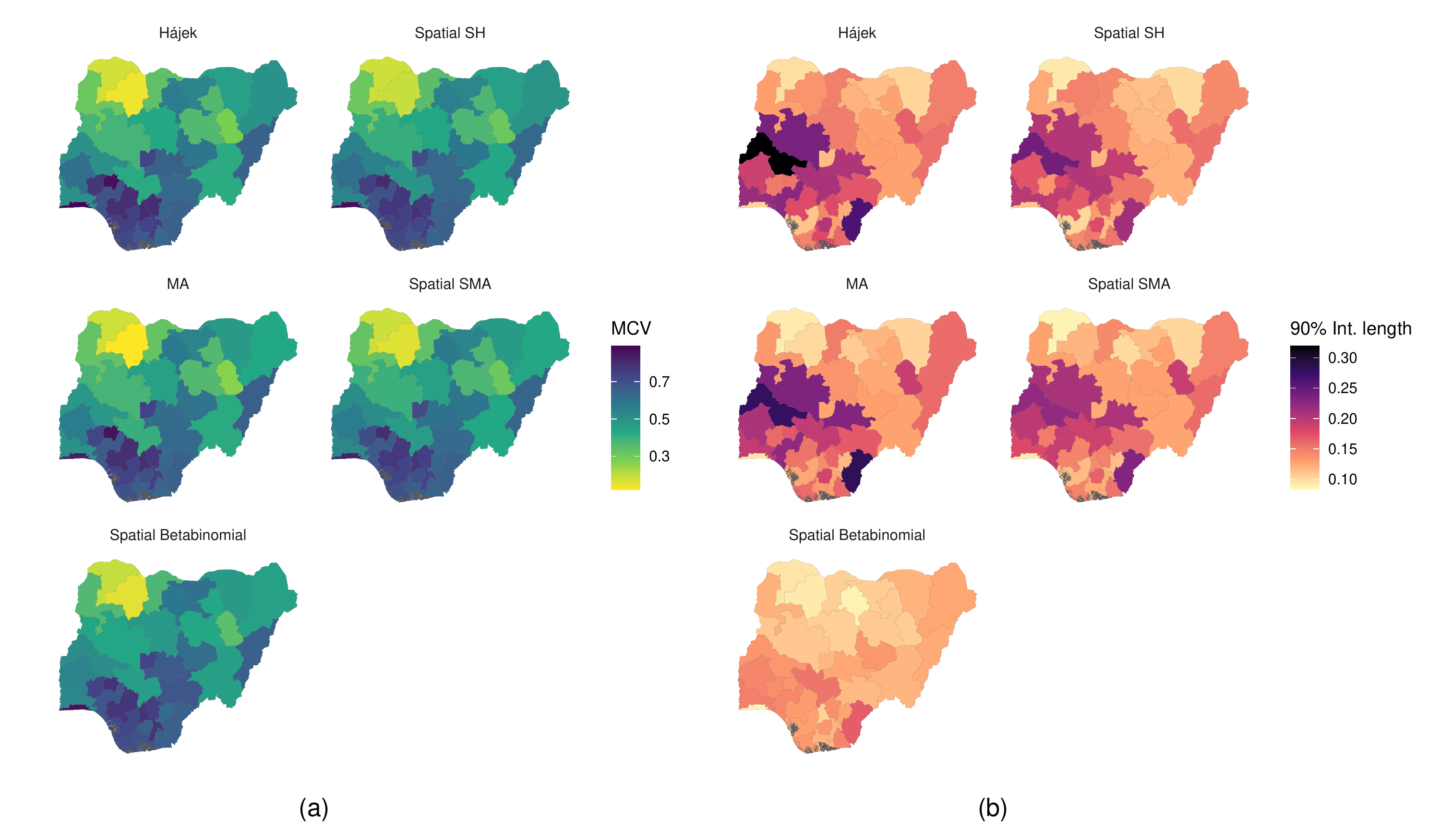}
	\caption{Estimated measles vaccination rates (left) and 90\% prediction interval lengths for estimated measles vaccination rates (right) among children aged 12-23 months for Admin-1 areas in Nigeria in 2018.}
	\label{fig:mcv-ests}
\end{figure}

Figure \ref{fig:mcv-ests} compares point estimates of measles vaccination rates (left) and the length of interval estimates (right) for Admin-1 areas among children aged 12-23 months in Nigeria in 2018. Point and interval estimates for all methods are provided in Appendix \ref{app:nga}. We omit results for the non-spatial mixed models as their results are similar to those of the spatial models. On the right side, we quantify uncertainty using the length of 90\% credible intervals (for the smoothed and unit level models) and design-based confidence intervals (for the H\'ajek and model-assisted estimators) for Admin-1 areas in Nigeria in 2018. The interpretation of uncertainty estimates requires some care since the intervals for the H\'ajek and model-assisted estimators only estimate design-based uncertainty, while the smoothed and betabinomial intervals are drawn from posterior distributions which also account for model parameter uncertainty. Although the point estimates for all the methods are similar, the interval estimate lengths vary considerably. In particular, incorporating unit level covariates shrinks the interval estimates as seen when comparing the H\'ajek and model-assisted estimators. Similarly, applying a smoothing model reduces estimated uncertainty; the interval lengths are shortest for the unit level model. Detailed estimates for these selected models are provided in the Appendix.
\begin{figure}
	\centering
	\includegraphics[scale=.43]{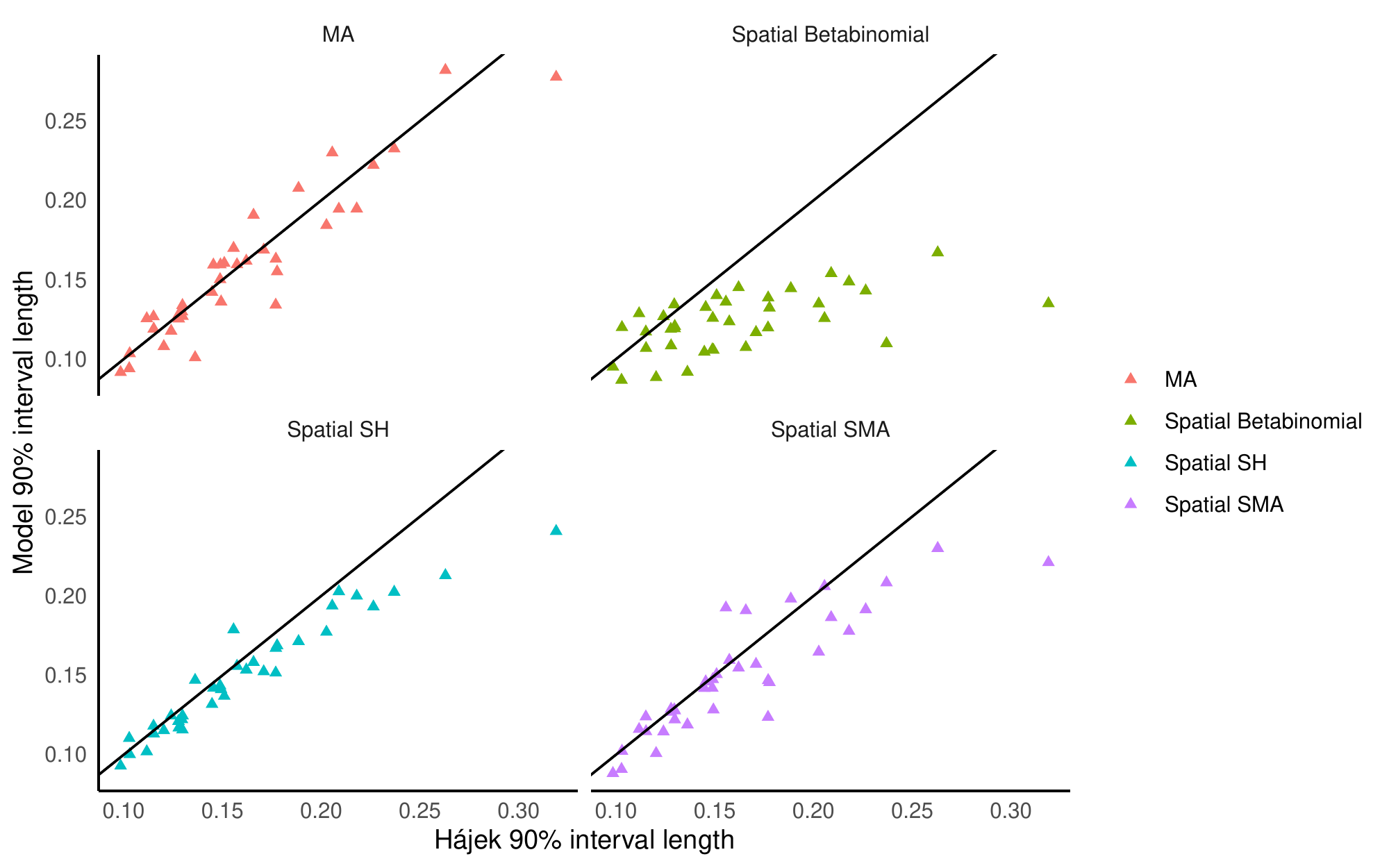}
	\caption{Estimated measles vaccination rates (left) and 90\% prediction interval lengths for estimated measles vaccination rates (right) among children aged 12-23 months for Admin-1 areas in Nigeria in 2018.}
	\label{fig:mcv-intlens}
\end{figure}

Figure \ref{fig:mcv-intlens} compares the interval lengths for the H\'ajek estimates with the lengths of interval estimates produced by the other methods, illustrating that the betabinomial intervals are considerably shorter than those produced by the rest of the methods. In particular, our smoothed model-assisted intervals are more conservative; as our simulations show, when relevant design variables are omitted from unit level models, resulting prediction intervals can exhibit undercoverage and corresponding smoothed model-assisted intervals may be better calibrated. In DHS surveys, clusters are sampled with probability proportional to size, but the cluster sizes are generally not published. As such, cluster size may be a relevant design variable that we are unable to incorporate into unit level models. The smoothed model-assisted point estimates and interval estimates may thus be preferable to the unit level model estimates. Among the unit level models, we recommend the use of the betabinomial estimates which account for potential clustering effects.

\section{DISCUSSION}\label{s:disc}

We have proposed a smoothed model-assisted estimator for small area means that incorporates unit level covariate information and smoothing via random effects while retaining favorable design optimality properties. Our method seeks to bridge the small area estimation and model-based geostatistics literatures, drawing from and offering benefits to both perspectives.

The basic question of how best to estimate area specific quantities given limited data arises in many settings; in a sense, any subpopulation with limited sample data may be considered a small area. For example, multilevel regression and poststratification has been used to generate local estimates of opinion using survey data with high nonresponse or nonprobability sampling (Si, 2020). Our smoothed model-assisted approach is particularly tailored for estimating subnational health and demographic indicators. In this context, properties like asymptotic design unbiasedness and design consistency are high priorities for national statistics offices that create and distribute estimates. Using spatial and spatio-temporal smoothing and unit level covariate information in small area estimation may offer large benefits in areas with limited data. Finally, the household surveys used in this setting typically have low non-response rates, informative sampling weights, and geographic information.

Although the above simulations and application illustrate potential benefits of our approach, in some settings, the new method may offer limited improvement. When data is not available in every area for which estimates are desired, it is still computationally possible to sample from the smoothing model posterior for unsampled areas.  However, as they do not incorporate actual observations from the area in question, such estimates cannot meaningfully be called design-consistent. Typical unit level models enable predictions to be made for unsampled areas and when sampling is not informative, such predictions may be preferable to those that would result from our approach. Another limitation is that our estimator requires careful specification of the working model. When the model is overly flexible, typical approximations that ignore variability from model estimation will underestimate the variance of model-assisted estimators. Resulting smoothed model-assisted estimators may thus be over confident  Finally, when reliable population information is unavailable, as may be the case when estimating health and demographic indicators in LMIC, it is common to use satellite-derived covariate rasters and aggregate pixel level predictions to compute area level estimates. This aggregation process is affected by measurement error in the covariates, misalignment between population density maps and household locations, and the resolution of the pixel grid. The effects of aggregation on the resulting area level estimates are not well understood; Paige et al. (2020) consider some of the potential implications. 

When unit level covariates are strongly associated with a variable of interest, using covariate modeling in small area estimation offers accuracy and efficiency gains. However, unit level models do not generally produce design-consistent estimators. Various solutions have been proposed, including pseudo-likelihood and specifically pseudo-Bayesian methods, pairwise likelihood methods, and direct modeling of the sample distribution. Pseudo-likelihood methods are sensitive to scaling and pairwise likelihood estimation requires knowledge of pairwise sampling probabilities, while uncertainty quantification for pseudo-Bayesian approaches relies on applying ad hoc corrections. Direct modeling of the sample distribution may necessitate undesirable model assumptions. As such, more work is needed to understand how best to use unit level covariate modeling in a setting where design optimality properties are prioritized.

\appendix

\section{Survey Asymptotics\label{app:asym}}

Let $\widehat{p}_a(s)$ denote an estimator of $p_a$ depending on the observed sample $s$. The expectation with respect to sampling $E_d$ can be defined as
\begin{equation} E_d(\widehat{p}_a(s))=\sum_s P_d(s)\widehat{p}_a(s)\end{equation}
and design variance $V_d$ can be analogously defined. 

In the survey statistics literature, it is also common to define survey asymptotics in terms of a sequence of nested samples and populations that are both increasing in size (S\"arndal, Swensson, and Wretman, 2003; Breidt and Opsomer, 2017). Let $U^{(\infty)} =1,2,\ldots$ be an infinite sequence of elements with associated $y$ values $y_1,y_2,\ldots$ and $U^{(1)},U^{(2)},\ldots$ be a sequence of populations where $U^{(k)}$ contains the first $N^{(k)}$ elements of $U^{(\infty)}$ and $U^{(1)}\subset U^{(2)}\subset\cdots$. For each $U^{(k)}$, let $P_d^{(k)}(\cdot)$ be a sampling design that assigns probabilities to each possible sample $s^{(k)}$. Assume sample size $n^{(k)}$ is fixed and $n^{(1)}<n^{(2)}<\cdots$. Thus $k\rightarrow\infty$ implies $n^{(k)}\rightarrow \infty$ and $N^{(k)}\rightarrow\infty$.
Let $\theta^{(k)}$ be a function of the elements of $U^{(k)}$ and let $\widehat{\theta}^{(k)}$ be an estimator of $\theta^{(k)}$ based on the sample $s_v$. An estimator $\widehat{\theta}^{(k)}$ is asymptotically design unbiased for $\theta^{(k)}$ if 
\begin{equation}\lim_{k\rightarrow\infty}[E_{d}^{(k)}(\widehat{\theta}^{(k)})-\theta^{(k)}]=0\end{equation}
and $\widehat{\theta}^{(k)}$ is design-consistent if for any fixed $\epsilon>0$,
\begin{equation}\lim_{k\rightarrow\infty}P_d^{(k)}(|\widehat{\theta}^{(k)}-\theta^{(k)}|>\epsilon)=0\end{equation}
S\"arndal et al. (2003) note that these conditions depend on the specification of the sequences of estimators, population values $\{y_i\}$, and designs $\{P^{(k)}\}$; conditions on the limiting behavior of the finite population values and inclusion probabilities are typically needed to ensure consistency.

\section{Design consistency of survey regression LGREG estimator\label{a:proof}}

We now consider the design consistency of the model-assisted estimator specified by Equation (\ref{e:greg-h}) and discuss the relevant regularity assumptions. Our proof adapts the one presented by Kennel and Valliant (Appendix, 2020) for a multivariate logistic model-assisted estimator for clustered samples. Rather than showing design consistency for $\widehat{p}_a^{MA}$, we instead consider the area-specific total estimator $\widehat{t}_a^{MA}$:
\begin{equation}\label{e:greg-total}
\widehat{t}_a^{MA}=\sum_{i\in U_a}\widehat{y}_i+\sum_{i\in S_a}w_i(y_i-\widehat{y}_i)
\end{equation}
Let $\widehat{y}_i$ denote predictions from our working logistic regression model:
\begin{align}
P(y_i=1\mid \mathbf{z}_i, \boldsymbol\gamma)&=q_i\\
\mathrm{logit}(q_i) &= \mathbf{z}_i^T\boldsymbol\gamma
\end{align}
If we had full population data, we could estimate $\boldsymbol\gamma$ by maximizing the population log-likelihood to obtain finite population parameters $\mathbf{G}$:
\begin{equation}
\mathbf{G}=\arg\max_{\boldsymbol\gamma} \sum_{i\in U}\ell(y_i; \boldsymbol\gamma)
\end{equation}
Since we only have data for sampled units $i\in S$, in practice, we maximize the survey-weighted log-likelihood to obtain $\widehat{\mathbf{G}}$, an estimator of $\mathbf{G}$:
\begin{equation}
\widehat{\mathbf{G}}=\arg\max_{\boldsymbol\gamma} \sum_{i\in S}\frac{1}{\pi_i}\ell(y_i; \boldsymbol\gamma)
\end{equation}
To reflect the dependence of our predictions on the estimated regression parameters we introduce the following notation, letting $\widetilde{y}$ denote predictions if we observed the finite population parameters $\mathbf{G}$:
\begin{align}
\widehat{y}_i&=\mu(\mathbf{z}_i, \widehat{\mathbf{G}})\\
\widetilde{y}_i&=\mu(\mathbf{z}_i, \mathbf{G})
\end{align}
We assume an asymptotic regime with a fixed number of $A$ areas, where area $a$ has sample size $n_a$ and population size $N_a$. We let $N$ denote the overall population size and $n$ denote the overall sample size. We assume a sequence of designs and populations such that $N, N_a\rightarrow\infty$ and assume the following conditions:
\begin{enumerate}
	\item The regression parameter estimates satisfy $\widehat{\mathbf{G}}=\mathbf{G}+O_p(n^{-1/2})$. Moreover, $\mathbf{G}\rightarrow\boldsymbol\gamma$ as $N\rightarrow\infty$.
	\item For each area $a$, for each $i$, $|\frac{\partial\mu}{\partial\mathbf{t}}|\leq h(\mathbf{z}_i, \boldsymbol\gamma)$ for all $\mathbf{t}$ in a neighborhood centered on $\boldsymbol\gamma$ such that $\frac{1}{N_a}\sum_{i\in U_a}h(\mathbf{z}_i,\boldsymbol\gamma)=O(1)$.
	\item  For each area $a$, $\sum_{i\in S_a}w_i\widetilde{y}_i$ is design-consistent for $\sum_{i\in U_a}\widetilde{y}_i$ and $\sum_{i\in S_a}w_iy_i$ is design-consistent for $\sum_{i\in U_a}y_i$.
\end{enumerate}
Note that Assumption 1 requires that the same working model holds for all areas or alternatively, that the survey design calls for proportional sampling of all areas $a$. Assumption 3 requires that Horvitz-Thompson type estimators are design-consistent under the sequence of designs specified.

By Taylor's theorem, for all $i \in U$, there is some vector $\mathbf{G}^*_i$ such that 
\begin{equation}
\widehat{y}_i=\widetilde{y}_i+\left[\frac{\partial\mu}{\partial\mathbf{t}}\bigg|_{\mathbf{t}=\mathbf{G}^*_i}\right]^T\mathrm{vec}(\widehat{\mathbf{G}}-\mathbf{G})
\end{equation}
Here, $\frac{\partial\mu}{\partial\mathbf{t}}$ is a $(p+1)\times 1$ vector of the partial derivatives of $\mu$ with respect to the components of $\mathbf{t}$. By summing over all units $i\in U_a$ and dividing by the population size $N_a$, we obtain the following:
\begin{equation}
\frac{1}{N_a}\sum_{i\in U_a}\widehat{y}_i=\frac{1}{N_a}\sum_{i\in U_a}\widetilde{y}_i+\frac{1}{N_a}\sum_{i\in U_a}\left[\frac{\partial\mu}{\partial\mathbf{t}}\bigg|_{\mathbf{t}=\mathbf{G}^*_i}\right]^T\mathrm{vec}(\widehat{\mathbf{G}}-\mathbf{G})
\end{equation}
Under Assumptions 1 and 2, we have that 
\begin{equation}
\frac{1}{N_a}\sum_{i\in U_a}\widehat{y}_i-\frac{1}{N_a}\sum_{i\in U_a}\widetilde{y}_i=O_p(n^{-1/2})
\end{equation}
and
\begin{equation}
\frac{1}{N_a}\sum_{i\in S_a}w_i\widehat{y}_i-\frac{1}{N_a}\sum_{i\in S_a}w_i\widetilde{y}_i=O_p(n^{-1/2})
\end{equation}
implying that
\begin{equation}
\frac{1}{N_a}\left[\sum_{i\in U_a}\widehat{y}_i-\sum_{i\in S_a}w_i\widehat{y}_i\right]=\frac{1}{N_a}\left[\sum_{i\in U_a}\widetilde{y}_i-\sum_{i\in S_a}w_i\widetilde{y}_i\right]+O_p(n^{-1/2})\end{equation}
We can thus rewrite $\widehat{t}_a^{MA}$ as follows:
\begin{align}
\frac{1}{N_a}\widehat{t}_a^{MA}&=\frac{1}{N_a}\left[\sum_{i\in U_a}\widehat{y}_i+\sum_{i\in S_a}w_i(y_i-\widehat{y}_i)\right]\\
&=\frac{1}{N_a}\left[\sum_{i\in S_a}w_iy_i+\sum_{i\in U_a}\widehat{y}_i-\sum_{i\in S_a}w_i\widehat{y}_i\right]\\
&=\frac{1}{N_a}\left[\sum_{i\in S_a}w_iy_i+\sum_{i\in U_a}\widetilde{y}_i-\sum_{i\in S_a}w_i\widetilde{y}_i\right]+O_p(n^{-1/2})
\end{align}
Therefore, as long as Assumption 3 holds, 
$\widehat{t}_a^{MA}$ will converge to the desired population total $\sum_{i\in U_a}y_i$.

\section{Parameter estimation\label{a:est}}

The analyses below were carried out using the R programming language (R Core Team, 2021). The R \texttt{survey} package provides tools for analyzing survey data and calculating commonly used small area estimators (Lumley, 2004). We also use the R package \texttt{INLA} to conduct approximate Bayesian inference. The \texttt{tidyverse} (Wickham et al., 2014), \texttt{sf} (Pebesma et al., 2018), and \texttt{raster} (Hijmans et al., 2021) packages were used to process data. The \texttt{R} package \texttt{SUMMER} (Li et al., 2020) can be used to fit similar models and functions for smoothed model-assisted estimation are currently in development.

We compute the H\'ajek estimators for all areas using the R package \texttt{survey}, which also provides associated variance estimates.For the simulations and application, the working logistic regression models are fit via survey-weighted maximum likelihood using the R package \texttt{survey}. Based on the working model predictions, model-assisted estimators are computed for each area and associated variance estimates are calculated using Kennel and Valliant's (2010) with-replacement cluster sampling variance estimator.

The area level models described in the main text take as input a set of direct or model-assisted estimates for all areas with associated variance estimates. We adopt a fully Bayesian approach to estimation by assuming priors on model parameters and using \texttt{INLA} to approximate the posterior distributions for area level proportions $p_a$ for all $a=1,\ldots, A$. We generate predictions by repeatedly sampling from these posterior distributions, enabling us to produce point estimates (from the posterior medians) and interval estimates (by taking relevant quantiles of the posteriors). The uncertainty of the resulting estimates may be quantified either using posterior variance or by taking the length of interval estimates.

For all area level models, we place a flat prior on the area level model intercept and fixed effects, so $\pi(\boldsymbol\beta)\propto 1$. As described above, we use penalized complexity priors for the variance parameters, as implemented in \texttt{INLA}. For the non-spatial area-level models, we specify the prior for the area effect variance $\sigma_u^2$ such that $P(\sigma_u>5)=0.01$. For the spatial area-level models, we specify the prior for the area effect variance $\sigma_u^2$ such that $P(\sigma_u>5)=0.01$ and for the spatial correlation parameter $\phi$ such that $P(\phi>.5)=2/3$. We select these priors to be relatively flat.

The unit level models described in the main text take as input survey microdata with covariate information for each sampled individual. As with the area level models, we use a fully Bayesian approach implemented using \texttt{INLA}. In order to generate predictions, we require covariate information for all sampled and non-sampled individuals to enable us to generate predictions for all individuals in our population of interest. 

For all unit level models, we place a flat prior on the intercept and fixed effects, so $\pi(\boldsymbol\gamma)\propto 1$. As described above, we use penalized complexity priors for the variance parameters. For the non-spatial area-level models, we specify the prior for the area effect variance $\sigma_u^2$ such that $P(\sigma_u>5)=0.01$. For the spatial area-level models, we specify the prior for the area effect variance $\sigma_u^2$ such that $P(\sigma_u>5)=0.01$ and for the spatial correlation parameter $\phi$ such that $P(\phi>.5)=2/3$.

\section{Additional results\label{app:nga}}
\begin{table}
	\renewcommand{\arraystretch}{0.5}
	\scriptsize
	\centering
	\tiny
\begin{tabular}{lp{.2cm}lp{.2cm}lp{.2cm}lp{.2cm}lp{.2cm}l}

\hline
State&\multicolumn{2}{c}{H\'ajek}&\multicolumn{2}{c}{Sp. SH}&\multicolumn{2}{c}{MA}&\multicolumn{2}{c}{Spatial SMA}&\multicolumn{2}{c}{Sp. Betabinomial}\\
\hline
Lagos & 0.89 & (0.84, 0.94) & 0.83 & (0.78, 0.88) & 0.82 & (0.77, 0.86) & 0.87 & (0.81, 0.92) & 0.86 & (0.81, 0.9)\\
Ekiti & 0.87 & (0.79, 0.94) & 0.86 & (0.77, 0.94) & 0.79 & (0.68, 0.87) & 0.80 & (0.7, 0.88) & 0.78 & (0.71, 0.85)\\
Anambra & 0.80 & (0.73, 0.88) & 0.79 & (0.72, 0.86) & 0.77 & (0.69, 0.83) & 0.78 & (0.71, 0.84) & 0.78 & (0.72, 0.83)\\
Enugu & 0.80 & (0.71, 0.88) & 0.77 & (0.69, 0.85) & 0.75 & (0.67, 0.81) & 0.76 & (0.67, 0.84) & 0.75 & (0.67, 0.81)\\
Edo & 0.79 & (0.7, 0.88) & 0.79 & (0.71, 0.87) & 0.77 & (0.69, 0.84) & 0.76 & (0.67, 0.84) & 0.77 & (0.7, 0.84)\\
Osun & 0.77 & (0.69, 0.85) & 0.73 & (0.65, 0.81) & 0.71 & (0.63, 0.78) & 0.75 & (0.68, 0.81) & 0.73 & (0.66, 0.8)\\
Abia & 0.75 & (0.69, 0.82) & 0.74 & (0.68, 0.81) & 0.73 & (0.67, 0.79) & 0.74 & (0.68, 0.8) & 0.78 & (0.72, 0.84)\\
Delta & 0.75 & (0.69, 0.8) & 0.74 & (0.67, 0.8) & 0.73 & (0.67, 0.79) & 0.74 & (0.69, 0.79) & 0.72 & (0.65, 0.78)\\
Abuja & 0.73 & (0.68, 0.79) & 0.74 & (0.67, 0.8) & 0.71 & (0.65, 0.78) & 0.72 & (0.66, 0.77) & 0.73 & (0.67, 0.79)\\
Imo & 0.73 & (0.63, 0.84) & 0.70 & (0.61, 0.79) & 0.70 & (0.61, 0.78) & 0.73 & (0.64, 0.81) & 0.65 & (0.58, 0.72)\\
Bayelsa & 0.73 & (0.65, 0.8) & 0.70 & (0.62, 0.78) & 0.70 & (0.62, 0.77) & 0.73 & (0.65, 0.8) & 0.71 & (0.64, 0.77)\\
Ondo & 0.69 & (0.58, 0.8) & 0.67 & (0.56, 0.78) & 0.67 & (0.57, 0.76) & 0.69 & (0.58, 0.78) & 0.68 & (0.6, 0.75)\\
Rivers & 0.68 & (0.59, 0.77) & 0.66 & (0.59, 0.72) & 0.66 & (0.6, 0.72) & 0.69 & (0.61, 0.76) & 0.69 & (0.63, 0.75)\\
Cross River & 0.65 & (0.52, 0.78) & 0.65 & (0.51, 0.79) & 0.65 & (0.53, 0.76) & 0.65 & (0.54, 0.76) & 0.65 & (0.56, 0.73)\\
Adamawa & 0.65 & (0.57, 0.73) & 0.66 & (0.58, 0.74) & 0.63 & (0.55, 0.71) & 0.62 & (0.54, 0.7) & 0.65 & (0.58, 0.71)\\
Nassarawa & 0.65 & (0.54, 0.75) & 0.62 & (0.5, 0.73) & 0.60 & (0.5, 0.7) & 0.63 & (0.53, 0.72) & 0.67 & (0.6, 0.73)\\
Ebonyi & 0.63 & (0.57, 0.7) & 0.63 & (0.58, 0.69) & 0.63 & (0.58, 0.69) & 0.64 & (0.58, 0.7) & 0.61 & (0.55, 0.68)\\
Akwa Ibom & 0.63 & (0.55, 0.71) & 0.64 & (0.56, 0.72) & 0.64 & (0.56, 0.72) & 0.64 & (0.56, 0.71) & 0.65 & (0.58, 0.72)\\
Benue & 0.63 & (0.54, 0.71) & 0.62 & (0.54, 0.7) & 0.62 & (0.54, 0.7) & 0.63 & (0.55, 0.7) & 0.68 & (0.62, 0.74)\\
Oyo & 0.60 & (0.51, 0.7) & 0.57 & (0.46, 0.67) & 0.57 & (0.47, 0.67) & 0.61 & (0.52, 0.69) & 0.54 & (0.47, 0.62)\\
Plateau & 0.59 & (0.52, 0.65) & 0.60 & (0.53, 0.66) & 0.59 & (0.52, 0.65) & 0.58 & (0.52, 0.64) & 0.61 & (0.54, 0.67)\\
Kano & 0.56 & (0.5, 0.62) & 0.58 & (0.52, 0.63) & 0.57 & (0.52, 0.62) & 0.56 & (0.5, 0.61) & 0.58 & (0.54, 0.63)\\
Jigawa & 0.54 & (0.48, 0.6) & 0.54 & (0.49, 0.6) & 0.54 & (0.48, 0.6) & 0.53 & (0.48, 0.59) & 0.60 & (0.55, 0.66)\\
Kwara & 0.51 & (0.35, 0.67) & 0.48 & (0.34, 0.62) & 0.52 & (0.41, 0.63) & 0.55 & (0.43, 0.67) & 0.52 & (0.45, 0.58)\\
Ogun & 0.51 & (0.4, 0.62) & 0.50 & (0.4, 0.6) & 0.53 & (0.44, 0.62) & 0.55 & (0.45, 0.65) & 0.53 & (0.46, 0.6)\\
Borno & 0.49 & (0.42, 0.57) & 0.43 & (0.35, 0.51) & 0.43 & (0.36, 0.51) & 0.49 & (0.42, 0.56) & 0.45 & (0.39, 0.52)\\
Yobe & 0.45 & (0.4, 0.5) & 0.47 & (0.42, 0.52) & 0.47 & (0.42, 0.52) & 0.45 & (0.4, 0.5) & 0.47 & (0.41, 0.53)\\
Kaduna & 0.43 & (0.35, 0.5) & 0.45 & (0.38, 0.52) & 0.45 & (0.39, 0.52) & 0.43 & (0.36, 0.5) & 0.48 & (0.43, 0.53)\\
Kogi & 0.42 & (0.32, 0.53) & 0.41 & (0.31, 0.5) & 0.46 & (0.37, 0.56) & 0.49 & (0.39, 0.59) & 0.46 & (0.38, 0.54)\\
Taraba & 0.42 & (0.36, 0.48) & 0.42 & (0.36, 0.48) & 0.43 & (0.36, 0.49) & 0.43 & (0.37, 0.49) & 0.45 & (0.39, 0.51)\\
Niger & 0.39 & (0.27, 0.51) & 0.38 & (0.27, 0.5) & 0.40 & (0.3, 0.51) & 0.41 & (0.31, 0.51) & 0.44 & (0.38, 0.49)\\
Bauchi & 0.36 & (0.3, 0.43) & 0.36 & (0.3, 0.42) & 0.37 & (0.31, 0.44) & 0.37 & (0.32, 0.43) & 0.43 & (0.38, 0.49)\\
Katsina & 0.34 & (0.26, 0.41) & 0.32 & (0.25, 0.4) & 0.34 & (0.26, 0.41) & 0.35 & (0.28, 0.43) & 0.37 & (0.32, 0.43)\\
Kebbi & 0.31 & (0.25, 0.38) & 0.33 & (0.26, 0.39) & 0.33 & (0.27, 0.39) & 0.31 & (0.25, 0.37) & 0.37 & (0.31, 0.43)\\
Gombe & 0.28 & (0.2, 0.36) & 0.26 & (0.17, 0.36) & 0.31 & (0.22, 0.41) & 0.31 & (0.24, 0.4) & 0.35 & (0.29, 0.4)\\
Sokoto & 0.18 & (0.13, 0.23) & 0.18 & (0.13, 0.22) & 0.18 & (0.14, 0.23) & 0.18 & (0.14, 0.23) & 0.19 & (0.15, 0.24)\\
Zamfara & 0.14 & (0.07, 0.21) & 0.12 & (0.07, 0.17) & 0.16 & (0.11, 0.23) & 0.19 & (0.13, 0.28) & 0.16 & (0.12, 0.21)\\
\hline
\end{tabular}

	\caption{Estimated measles vaccination rates (left) with 90\% prediction intervals for Admin-1 areas in Nigeria in 2018.}
	\label{fig:mcv-ests-table}
\end{table}
\end{document}